\documentclass{article}

\usepackage{amssymb,amsmath,latexsym}

\usepackage{graphicx}

\newtheorem{theorem}{Theorem}[section]
\newtheorem{algorithm}[theorem]{Algorithm}

\newtheorem{corollary}[theorem]{Corollary}

\newtheorem{definition}[theorem]{Definition}
\newtheorem{example}[theorem]{Example}

\newtheorem{proposition}[theorem]{Proposition}
\newtheorem{remark}[theorem]{Remark}

\title{On A. V. Anisimov's problem for finding a polynomial algorithm checking inclusion of context-free languages in group languages}
\author{Krasimir Yordzhev}
\date{}

\begin{document}

\maketitle

Trakia University, Stara Zagora, Yambol, Bulgaria

Email address: krasimir.yordzhev@gmail.com 

\begin{abstract}
The work investigates the problem of whether a context-free language is a subset of a group language. A.~V. Anisimov has shown that the problem of determining the unambiguity of finite automata is a special case of this problem. Then the question of finding polynomial algorithm verifying the inclusion of context-free
languages in group languages naturally arises.   The article focuses on this open problem. For the purpose, the paper describes an unconventional method of description of context-free languages, namely a representation with the help of a finite digraph whose arcs are labelled with a specially defined monoid $\mathcal{U}$. Also, we define a semiring $\mathcal{S}_\mathcal{U}$ whose elements are the set $2^\mathcal{U}$ of all subsets of $\mathcal{U}$ and with operations - product and union of the elements of $2^\mathcal{U}$. The described algorithm executes no more than $O(n^3)$ operations in $\mathcal{S}_\mathcal{U}$.
\end{abstract}

\section{Introduction}

The work is a continuation and significant improvement of the results obtained in the publication \cite{RN63}.

Let $G$ be a group with the identity $e$ and with the set of generators
\begin{equation}\label{SigmaGroup}
\Sigma=X \cup X' = \left\{ x_1 ,x_2,...,x_m \right\} \cup \left\{ x_1' ,x_2' ,...,x_m' \right\} ,\quad X \cap X' =\emptyset
\end{equation}
and the set of defining  relations $\Theta$ such that
\begin{equation}\label{ThetaGroup}
\left\{ x_i x_i' =  x_i' x_i  =e \; |\; i=1,2,\ldots ,m \right\} \subseteq \Theta .
\end{equation}

\begin{definition}
If
$$
\mathfrak{L} (G) =\left\{ \left. \omega \in \Sigma^* \; \right\vert \;\omega \equiv e \; (\textrm{mod} \; G) \right\} \subseteq \Sigma^* ,
$$
then $\mathfrak{L} (G)$ we will call a \textbf{group language} representing $G$, where $\Sigma^*$ is a free monoid over $\Sigma$ and $e$ is the identity in $G$.
\end{definition}

A. V. Anisimov introduces the concept of group language in \cite{springerlink:10.1007/BF01071030}. In just cited article, Anisimov proved that $\mathfrak{L} (G)$ is regular if and only if the group $G$ is finite (See also   \cite[Theorem 5.17]{Chiswell}).

A somewhat different definition of the concept of group language is given in \cite{Heam20115808}, namely a regular language whose syntactic monoid is a finite group. In our work, we will stick to the first definition given by A.V. Anisimov.

In \cite{springerlink:10.1007/BF01068773} A.~V. Anisimov has showed that the problem of determining the unambiguity of finite automata is a special case of the problem of determining whether a context-free language is a subset of a group language. Then the problem of finding polynomial algorithms verifying the inclusion of context-free
languages in group languages naturally arises. This problem is solved in its particular cases for regular and linear languages (which are  special cases of context-free languages) in \cite{RN63}, where it is shown that the inclusion of a regular or a linear language in a group language can be decided in polynomial time.

In \cite{springerlink:10.1007/BF01068773} A.~V. Anisimov gives an algorithm to check whether the inclusion $L \subseteq \mathfrak{L}(G)$ is true. Unfortunately, this algorithm is not polynomial.

The aim of the present work is to describe a polynomial algorithm that solves the problem formulated by A.~V. Anisimov for an arbitrary context-free language.

\section{Preliminaries}

Let $\Sigma$ be a finite and non-empty set, which we will call {\it alphabet}. The elements of this set we will call {\it letters}.
We will call a \emph{word over the alphabet} $\Sigma$ each finite string of letters from $\Sigma$.
 A word that does not contain any letter is called an \emph{empty word}, which we will mark with $\varepsilon$. $\Sigma^*$ denotes the \emph{free monoid} with the identity $\varepsilon$, i.e. the set of all words over $\Sigma$, including empty set with operation \emph{concatenation}. $\Sigma^+ =\Sigma^* \setminus \{ \varepsilon \}$. The term \emph{ length of a word} refers to the number of letters in it. The length of the word $\alpha$ will be expressed by $|\alpha |$. By definition $|\varepsilon |=0$.  Each subset  $L\subseteq \Sigma^*$ is called \emph{ formal language} (or only \emph{ language})  over alphabet $\Sigma$. 

According to \cite{Lallement} a \emph{context-free grammar} $\Gamma$ we will call the triple $\Gamma = \langle \mathcal{N} ,\Sigma ,\Pi \rangle$, where $\mathcal{N}$, $\Sigma$  are finite sets of nonterminals and terminals, respectively, $\mathcal{N} \cap \Sigma =\emptyset$ and $\Pi$ is a finite subset of the Cartesian product $\mathcal{N}\times \left( \mathcal{N} \cup \Sigma \right)^*$, whose elements are called \emph{productions} or \emph{rules}. The elements of $\Pi$ are denoted $A\to\omega$, where $A\in \mathcal{N}$, $\omega\in (\mathcal{N}\cup \Sigma )^*$. The notation $A\Rightarrow \omega$ indicates that there exists a sequence $A\to \alpha_1 A_1 \beta_1$, $A_1 \to \alpha_2 A_2 \beta_2$, $\ldots$ , $A_{t-2} \to \alpha_{t-1} A_{t-1} \beta_{t-1}$, $A_{t-1} \to \gamma$, where $A_i \in \mathcal{N}$ and $\alpha_i , \beta_i \in \left( \mathcal{N} \cup \Sigma \right)^*$ for every $i=1,2,\ldots t-1$, $\gamma \in \left( \mathcal{N} \cup \Sigma \right)^*$ and $\omega = \alpha_1 \alpha_2 \cdots \alpha_{t-1} \gamma \beta_{t-1} \beta_{t-2} \cdots \beta_1$. This sequence is called a \emph{derivation} with \emph{length} t.

Let $\Gamma = \langle \mathcal{N} ,\Sigma ,\Pi \rangle$ be a context-free  grammar an let $A\in \mathcal{N}$. Then the set $L(\Gamma, A) = \left\{ \alpha \in \Sigma^* \; \; |\; A\Rightarrow \alpha \right\}$ is the \emph{context--free language} generated by the grammar $\Gamma$ with the \emph{starting symbol} $A$. 

Throughout this article, we will assume that every nonterminal symbol $A\in \mathcal{N}$ is essential, i.e. $L(\Gamma ,A)\ne \emptyset$ for every $A\in \mathcal{N}$.


It is well known  \cite{dshtr,Rayward-Smith,syt} that any context-free language can be generated by some grammar in \emph{Chomsky normal form}, i.e. a grammar in which all the productions have the form $A\to BC$ or $A\to a$, where $A,B,C\in\mathcal{N}$ are nonterminals and $a\in\Sigma$ is a terminal.

Let $M$ be a finitely generated monoid with the set of generators $\Sigma$, the set of defining  relations $\Psi$, unit element $e$ and with decidable  word problem. Then the set of words
\begin{equation}\label{L(G)}
\mathfrak{L} (M) =\left\{ \left. \omega = a_{i_1} a_{i_2} \ldots a_{i_k} \in \Sigma^* \; \right\vert \; \omega = e\ {\rm is\ satisfied\ in}\ M  \right\}
\end{equation}
we will  call a \emph{monoidal language}, which specifies the monoid $M$.
The monoid $M$ is  specified by a context-free language, if the relevant monoidal language  $\mathfrak{L} (M)$ is  context-free. The monoid $M$ in this case is called a \emph{context-free monoid}.

In the case, that the monoid $M$ has  the set of generators (\ref{SigmaGroup}) and the set of defining  relations
\begin{equation}\label{ThetaDyck}
\Psi =\left\{ x_i x_i ' =e \; |\; i=1,2,\ldots ,n \right\},
\end{equation}
then $\mathfrak{L} (M)$ is called \emph{restricted Dyck language on the $2n$ letters from $\Sigma$}, which we will denote by $\mathfrak{D}_{2n}$. In this case, $x_i$ is called an \emph{opening bracket} and $x_i '$ is the corresponding \emph{closing bracket}.

For more information on automata and language theory we refer the reader to \cite{aho_ulman,ginzburg,Hopcroft}. For the mathematical foundations and algebraic approach of formal language theory we refer to \cite{Lallement,PerrinPin}. For the connections between formal language theory and group theory we recommend the source   \cite{Chiswell}.  A list of problems related to the discussed in this paper topics is given in \cite{Grigorchuk}.

Let $L$ be a context-free language and let $p$ and $q$  be the constants of the pumping lemma ($xuwvy$-theorem) for $L$ (see \cite[Lemma 3.1.1]{ginzburg},  \cite[Theorem 7.18]{Hopcroft},  or \cite[Theorem 5.3]{Rayward-Smith} ).
We define the sets:
\begin{description}
\item[$\Omega_1$] = $\displaystyle \left\{ \omega\in L \; \left|\; |\omega |\le p \right. \right\}$;
\item[$\Omega_2$] = $\displaystyle  \left\{ \left. uwvw' \; \right\vert \; \vert uwv\vert \le q,\; uv\ne\varepsilon ,\; \exists A\in N : A\Rightarrow uAv, A\Rightarrow w  \right\} $;
\item[$W_1$] = $\Omega_1 \cup \Omega_2$.
\end{description}

The following theorem is proved in \cite{springerlink:10.1007/BF01068773}:

\begin{theorem}\label{anisimov}
{\rm (A. V. Anisimov  \cite{springerlink:10.1007/BF01068773}) }
Let $L$ be a context-free language and let $G$ be a group with the set of generators (\ref{SigmaGroup}), and the set of defining  relations $\Theta$ satisfying the condition (\ref{ThetaGroup}). Then $L \subseteq \mathfrak{L} (G)$ if and only if   $W_1 = \Omega_1 \cup \Omega_2 \subseteq \mathfrak{L} (G)$.

\hfill $\Box$
\end{theorem}

Theorem \ref{anisimov} gives an algorithm to check whether the inclusion $L \subseteq \mathfrak{L}(G)$ is true. Unfortunately, this algorithm is not polynomial. The works \cite{RN63} modify Anisimov's algorithm so that it works polynomially in special cases when $L$ is a regular or a linear language.

Recall that a \emph{directed graph} (or \emph{digraph} for short) $D$ is a pair $D = \langle V,R\rangle$ where $V$ is a nonempty set, and $R$ is a multiset of ordered pairs of elements from $V$. The elements of $V$ are the \emph{vertices} (or \emph{nodes}) of the digraph $D$, the elements of $R$ are its \emph{arcs} (or \emph{oriented edges}). An arc whose beginning coincides with its end is called a \emph{loop}.
A \emph{walk} of \emph{length} $t$ in a digraph $D = \langle V,R\rangle$ is a  sequence $\rho_1  \rho_2 \ldots \rho_{t} $ of arcs $\rho_i$,  such that $\rho_i \in R$, $i=1,2,\ldots t$ and the end of $\rho_i$ coincides with the begin of $\rho_{i+1}$, $i=1,2,\ldots t-1$. A walk whose beginning coincides with its end is called a cycle.

For more details on graph theory see \cite{diestel,harary} for example.

The widespread use of graph theory in different areas of science and technology is well known. For example, graph theory is a good tool for the modelling of computing devices and computational processes and in some non-traditional areas, such as social science or modelling some processes in education and other humanitarian activities \cite{harary1953graph,OrozKrAt,OrAtTo}. So, many of graph algorithms have been developed \cite{swami}.

A \emph{transition diagram} is a 4-tuple $H=\langle V,R,\mathcal{S},l\rangle$, where $\langle V,R\rangle$ is a directed graph with the set of vertices $V$ and the multiset of arcs $R$; $\mathcal{S}$ is a semigroup whose elements will be called \emph{labels} and $l$ is a mapping from $R$ to $\mathcal{S}$, which we will call \emph{labeling mapping}.

If $\pi=p_1 \; p_2 \; \cdots \; p_k$ is a walk in H, $p_i \in R$, $i=1,2,\ldots k$ such that the end of $p_i$ coincides with the begin of $p_{i+1}$, $i=1,2,\ldots k-1$, then  $$l(p_1 \; p_2 \; \cdots \;p_k )=l(p_1 )l(p_2 )\ldots l(p_k ).$$

If $P$ is a set of walks in $H$, then
$$\displaystyle l(P )=\bigcup_{\pi\in P} l(\pi ) = \{\omega \in S\;\vert \; \exists \pi \in P : l(\pi )=\omega \} .$$



\section{A graph representation of context-free languages}\label{grphsct}

A classic example of the representation of context-free languages using finite digraphs is the transition diagram of pushdown automaton - recognizer of the corresponding context-free language. The paper \cite{RN77} describes a qualitatively new recognizer of context-free languages, based on some operations from graph theory. In the present article, we continue the work started in the mentioned above paper by improving the model and making it more user-friendly by adding new features and new useful tools.

\begin{definition}\label{WWW}
Let $\Sigma$ and $\cal N$ be finite sets, $\Sigma \cap \mathcal{N} =\emptyset$ and let $\Sigma^*$ be the free monoid over $\Sigma$ with the identity $\varepsilon$, where $\varepsilon$ is the empty word. We define the set
$$
  \mathcal{N}' = \left\{ A'\; |\; A\in \cal N \right\} ,\quad \cal N'\cap N =\emptyset .
$$

We define the monoid $T$ with the set of generators $\mathcal{N}\cup \mathcal{N}' \cup \{ e\}$, the identity $e$ and the set of defining relations

\begin{equation}\label{AA'}
  AA'=e ,\ Xe=eX=X, \quad  A\in \mathcal{N},\ A'\in \mathcal{N}' ,\ X\in \mathcal{N}\cup\mathcal{N}'.
\end{equation}

Let
$$
  \mathcal{U}=\Sigma^* \times T=\left\{ \langle \alpha ,\omega \rangle \; |\; \alpha\in \Sigma^* ,\; \omega\in T\right\} .
$$

In $\mathcal{U}$ we define the operation
\begin{equation}\label{circ}
  \langle \alpha_1 ,\omega_1 \rangle \circ \langle \alpha_2 ,\omega_2 \rangle =\langle \alpha_1 \alpha_2 ,\omega_1 \omega_2 \rangle ,
\end{equation}
where $\alpha_1 ,\alpha_2 \in \Sigma^*$, $\omega_1 ,\omega_2 \in T$. It is easy to see that $\mathcal{U}$ with the operation defined above is a monoid with unity element
\begin{equation}\label{unitofmonU}
1_\mathcal{U} = \langle \varepsilon ,e \rangle.
\end{equation}
\end{definition}

Obviously if $\omega\in T$, then

$$
\omega =e  \Longleftrightarrow \omega\in \mathfrak{D}_{2n} ,
$$
where $n=| \mathcal{N} |$ and $\mathfrak{D}_{2n}$ is  restricted Dyck language on the $2n$ letters from $\mathcal{N}\cup \mathcal{N}'$.

\begin{definition}\label{H_Gamma}
Let $\Gamma =\langle \mathcal N, \Sigma ,\Pi \rangle$ be a grammar in Chomsky normal form. Let $\mathcal{U}$  be the monoid defined by Definition \ref{WWW}. We construct the transition diagram
$$
  H_\Gamma =\langle V, R,\mathcal{U},l\rangle
$$
with the \emph{set of vertices}
$$
  V=\mathcal{N} \cup \{ Z\},\quad Z\notin \mathcal{N}
$$
and the \emph{multiset of arcs}
$$
  R\subseteq \left\{ \overrightarrow{AB}\; |\; A,B\in V \right\} .
$$

We label the arcs of $H_\Gamma$ using the function
$$
  l : R \to \left\{ \langle a, e \rangle \; | \; a\in \Sigma  \right\} \cup \left\{ \langle \varepsilon , Y\rangle \; | \; Y\in \mathcal{N}\cup \mathcal{N}' \right\} \subset \mathcal{U}.
$$

Each arc in $H_\Gamma$ satisfies one of the following conditions:
\begin{description}
  \item[(a)]\label{usl1} For every  production $A\to a \in \Pi$, where $A\in \mathcal{N}$ and $a\in\Sigma \cup \{ \varepsilon \}$,  there is an arc $\overrightarrow{AZ} \in R$ labeled $$l(\overrightarrow{AZ}) = \langle a,e\rangle;$$

  \item[(b)]\label{usl2} For every  production $A\to BC\in \Pi$, where $A,B,C\in \mathcal{N}$, there are arcs $\overrightarrow{AB} \in R$ and $\overrightarrow{ZC} \in R$ with labels respectively $$l(\overrightarrow{AB}) = \langle \varepsilon ,C\rangle \ \textrm{and} \  l(\overrightarrow{ZC}) = \langle \varepsilon ,C' \rangle;$$

\item[(c)]  There are no other arcs in $H_\Gamma$ except described in conditions (a) and  (b).
\end{description}
\end{definition}

\begin{theorem}\label{th2}
Let $\Gamma =\langle \mathcal N, \Sigma ,\Pi \rangle$ be a grammar in Chomsky normal form and let $H_\Gamma$ be the transition diagram obtained according to Definition \ref{H_Gamma}. Let $A\in \mathcal{N}$, $\alpha \in \Sigma^*$. Then $\alpha\in L(\Gamma , A)$ if an only if there is a walk $\pi$ with begin vertex $A$, end vertex $Z$ ($Z\notin \mathcal{N}$) and having label $l(\pi )=\langle \alpha , \omega \rangle =\langle \alpha , e \rangle$, where $e$ is the identity of monoid $T$ defined in Definition \ref{WWW}, equation (\ref{AA'}), i.e.  $\omega \in \mathfrak{D}_{2n}$, where $\mathfrak{D}_{2n}$ is restricted Dyck language on the $2n$ letters from $\mathcal{N}\cup \mathcal{N}'$, $n=|\mathcal{N} |=|\mathcal{N}' |$.
\end{theorem}

Proof.
Necessity. Let $A\in \mathcal{N}$ and let $\alpha \in L(\Gamma ,A)$. Then there is a derivation $A\Rightarrow \alpha$. Let the length of this derivation be equal to $t\ge 1$. We will prove the necessity by induction on $t$.

Let $t=1$. Since $\Gamma$ is a grammar in Chomsky normal form, then $\alpha =a$, where $a\in \Sigma \cup \{ \varepsilon \}$, and $A\to a$ is a production from $\Gamma$. According to condition (a) in Definition \ref{H_Gamma}, in $H_\Gamma$ there is an arc $\overrightarrow{AZ}$ with label $l(\overrightarrow{AZ} )=\langle a,e \rangle =\langle \alpha ,e \rangle$. Therefore, when $t = 1$ the necessity is fulfilled.

Suppose that for all $A\in \mathcal{N}$ and for all $\alpha \in L(\Gamma, A)$ for which there is a derivation $A\Rightarrow \alpha$ with length not greater than $t$, in $H_\Gamma$ there is a walk  with the start vertex $A$, the final vertex $Z$ and having label $\langle \alpha , \omega \rangle =\langle \alpha , e \rangle$, where $\omega \in \mathfrak{D}_{2n}$.

Let $A\Rightarrow \alpha$ is a derivation in $\Gamma$  which length is equal to $t+1$ and let $A\to BC$, $A,B,C\in \mathcal{N}$ be the first production in this derivation. Then in $\Gamma$ there exist derivations $B\Rightarrow \alpha_1$ and $C\Rightarrow \alpha_2$ with lengths not greater than $t$, where $\alpha_1 ,\alpha_2 \in \Sigma^*$ and $\alpha_1 \alpha_2 =\alpha$. By the inductive assumption, in $H_\Gamma$ there are:

i) a walk $\pi_1$ with the start vertex $B$, final vertex $Z$, labeled $l(\pi_1 )=\langle \alpha_1 ,e \rangle$ and

ii) a walk $\pi_2$ with start vertex $C$, final vertex $Z$ and labeled $l(\pi_2 )=\langle \alpha_2 ,e \rangle$.

According to Definition \ref{H_Gamma}, condition (b), in $H_\Gamma$ there are arcs $\overrightarrow{AB}$ and $\overrightarrow{ZC}$ with labels $l(\overrightarrow{AB} )=\langle \varepsilon ,C\rangle$ and $l (\overrightarrow{ZC} )=\langle \varepsilon , C' \rangle$ respectively. Then the walk $\pi =\overrightarrow{AB}\pi_1 \overrightarrow{ZC}\pi_2$ has  start vertex $A$, final vertex $Z$ and label:

$l(\pi )=l(\overrightarrow{AB} )\circ l(\pi_1 )\circ l(\overrightarrow{ZC})\circ l(\pi_2 )=
\langle \varepsilon ,C\rangle \circ \langle \alpha_1 ,e \rangle\circ \langle \varepsilon ,C' \rangle\circ \langle \alpha_2 ,e \rangle =
\langle \varepsilon\alpha_1 \varepsilon\alpha_2 , C e C' e \rangle =\langle \alpha_1 \alpha_2 ,CC' \rangle =\langle \alpha ,e \rangle .$

This proves the necessity.

Sufficiency. Let $A\in\mathcal{N}$ and let in $H_\Gamma$ there is a walk $\pi$ with start vertex $A\in \mathcal{N}$, final vertex $Z\notin \mathcal{N}$ an label  $l(\pi)=\langle \alpha , \omega \rangle =\langle \alpha ,e \rangle$, where $\alpha\in\Sigma^* $, $\omega \in \mathfrak{D}_{2n}$. We will prove the sufficiency by induction on the length $|\alpha |$ of the word $\alpha$.

If $|\alpha |=0$ or $|\alpha |=1$, then $\alpha =a$ for some $a\in \Sigma \cup \{ \varepsilon \}$. Hence $\pi = \overrightarrow{AZ}$ (see Definition \ref{H_Gamma}) and $\pi$ is an arc with label $l(\pi )=\langle a,e \rangle$.  Then according to Definition \ref{H_Gamma}, condition (a), in $\Gamma$ there is a production $A\to a$, i.e. $\alpha=a\in L(\Gamma , A)$.

Let $t$ is a positive integer, such that for every vertex $A\in \mathcal{N}$ and every walk $\pi$ in $H_\Gamma$ with start vertex $A$,  final vertex $Z$ and label $l(\pi )=\langle \alpha , \omega \rangle =\langle \alpha ,e \rangle$, $\alpha \in \Sigma^*$, $\omega \in \mathfrak{D}_{2n}$, from $|\alpha |\le t$ follows $\alpha\in L(\Gamma , A)$.

\begin{figure}[!h]
\begin{center}
\includegraphics{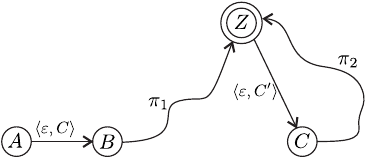}
\caption{}\label{fg1}
\end{center}
\end{figure}

Let $\alpha\in\Sigma^+$, where $|\alpha | = t+1\ge2$ and let $\pi$ be a walk in $H_\Gamma$ with start vertex $A\in \mathcal{N}$, final vertex $Z$ and label $l(\pi ) = \langle \alpha ,\omega \rangle=\langle \alpha ,e \rangle$, where $\omega \in \mathfrak{D}_{2n}$, and therefore there is not $A' \in\mathcal{N}'$ such that $A'$ is the first letter of $\omega$. Since $|\alpha |\ge 2$, there exists a vertex $B\in\mathcal{N}$ (i.e. $B\neq Z$), such that the first arc of $\pi$ is $\overrightarrow{AB}$ and let $l(\overrightarrow{AB} )=\langle \varepsilon ,C\rangle$, where $C\in\mathcal{N}$. But $l(\pi )=\langle \alpha ,e  \rangle$. Therefore, in order for the letter $C$ to disappear from the label of $\pi$, it follows that in $H_\Gamma$ there exist an arc $\overrightarrow{ZC}$ with the label $\langle \varepsilon ,C' \rangle$ and walks  $\pi_1$ and $\pi_2$, where $\pi_1$ has  start vertex $B$, final vertex $Z$ and $\pi_2$ has start vertex $C$, final vertex $Z$, such that the path $\pi$ is represented in the form $\pi = \overrightarrow{AB} \pi_1 \overrightarrow{ZC} \pi_2$ (see Figure \ref{fg1}).

Let $$l(\pi_1 )=\langle \alpha_1 ,\omega_1 \rangle ,\quad l(\pi_2 )=\langle \alpha_2 ,\omega_2 \rangle,$$ where $\alpha_1 ,\alpha_2 \in \Sigma^*$ and $\omega_1 ,\omega_1 \in (\mathcal{N}\cup\mathcal{N}' )^*$. Then we get:
$$l(\pi )=l(\overrightarrow{AB} \pi_1 \overrightarrow{ZC} \pi_2 )= l(\overrightarrow{AB} )\circ l( \pi_1 )\circ l(\overrightarrow{ZC} )\circ l(\pi_2 )=$$
$$=\langle \varepsilon ,C\rangle \circ \langle \alpha_1 ,\omega_1 \rangle \circ \langle \varepsilon ,C' \rangle \circ \langle \alpha_2 ,\omega_2 \rangle = $$
$$=\langle \alpha_1 \alpha_2 ,C\omega_1 C' \omega_2 \rangle .$$

Without loss of generality, we can assume that the vertex $C$ is not contained inside the walk $\pi_2$ and therefore $C'\notin \omega_2$.
From $l(\pi )=\langle \alpha , \omega \rangle$ we obtain $\alpha_1 \alpha_2 =\alpha$ and $\omega = C\omega_1 C'\omega_2 \in \mathfrak{D}_{2n}$. As $\omega \in \mathfrak{D}_{2n}$ and $C'\notin \omega_2$, then  $\omega_1$ is enclosed by the pair of opening bracket $C$ and corresponding closing bracket $C'$. Then it is easy to see that  $\omega_1 =e$ and therefore $\omega_2 =e$. Since $|\alpha_1 |\ge 1$, $|\alpha_2 |\ge 1$ and $|\alpha_1 |+|\alpha_2 |=|\alpha |$, we have $|\alpha_1 | < |\alpha |=t+1$ and $|\alpha_2 | < |\alpha |=t+1$, i.e. $|\alpha_1 |\le t$ and $\alpha_2 \le t$. By the inductive hypothesis, $\alpha_1 \in L(\Gamma ,B)$ and $\alpha_2 \in L(\Gamma ,C)$, i.e. in $\Gamma$ there exist derivations $B \Rightarrow \alpha_1$ and $C \Rightarrow  \alpha_2$. Therefore in $\Gamma$ there is a derivation $A\to BC \Rightarrow \alpha_1 C  \Rightarrow \alpha_1 \alpha_2 =\alpha$. This proves the sufficiency.

\hfill $\Box$

\begin{example}\rm
Consider the context-free grammar in Chomsky normal form $\Gamma =\langle \{ S,A,B,C,D\}$, $\{ a,b\} $, $\{ S\to SS$, $S\to AB$, $S\to BA$, $S\to AD$, $S\to BC$, $C\to SA$, $D\to SB$, $A\to a$, $B\to b\} \rangle$. It is easy to prove that $L(\Gamma , S)$ is the language of all words in $\{ a,b\}^* \setminus \{ \varepsilon \}$ in which the number of letters ''$a$'' is equal to the number of letters ''$b$''.
 The corresponding graph $H_\Gamma$ is shown in Figure \ref{fg2}.

\begin{figure}[!h]
\begin{center}
\includegraphics{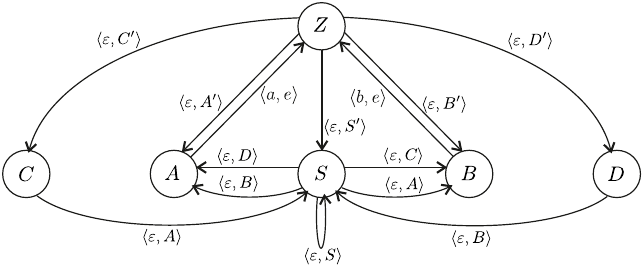}
\caption{}\label{fg2}
\end{center}
\end{figure}

\end{example}

From Theorem \ref{th2} follows the next theorem formulated and proved by Chomsky and Sch\"{u}tzenberger in \cite{CHOMSKY1963118}.

\begin{theorem} \label{Chomskyth}
{\rm \cite{CHOMSKY1963118} (See also   \cite[Theorem 5.14]{Lallement} or  \cite[Theorem 11.9]{Pentus})}
A language $L\subseteq \Sigma^*$ is context-free if and only if there are a positive integer $n$, a regular language $L_1$ over the alphabet $T= {\cal N}\cup {\cal N}'$, $|{\cal N}|=n$ and ${\cal N}' =\{ A' \; |\;  A\in \mathcal{N} \}$ and homomorphism $h : T^* \to \Sigma^*$ such that $L= h( \mathfrak{D}_{2n} \cap L_1 )$, where  $ \mathfrak{D}_{2n}$ is the restricted Dyck language on the $2n$ letters from the set $T$.

\hfill $\Box$
\end{theorem}


\section{Inclusion of context-free  languages in group languages}\label{section4}

Let $\cal N$, $\cal N'$, $X$, $X'$ and $\Sigma$ be finite sets, where
$$ \mathcal{N}=\{ A_1 , A_2 ,\ldots ,A_n \} ,\quad \mathcal{N}'=\{ A'_1 , A'_2 ,\ldots ,A'_n \} , \quad \mathcal{N}\cap \mathcal{N}' =\emptyset ,$$
$$ X= \left\{ x_1 ,x_2,...,x_m \right\} ,\quad X' = \left\{ x_1' ,x_2' ,...,x_m' \right\} ,\quad X \cap X' =\emptyset $$
and
$$
\Sigma=X \cup X' .
$$

Let
$$
a\in\mathcal{N} \cup\mathcal{N}'\cup X \cup X' .
$$
Then  \emph{by definition}, we put:
$$
(a')' =a .
$$

If
$$
\alpha =y_1 ,y_2 ,\ldots , y_l \in (X\cup X' )^* \quad \textrm{and} \quad \omega =a_1 a_2 \ldots a_k \in (\mathcal{N} \cup \mathcal{N}')^*
$$
then \emph{by definition}
$$
\alpha' = y_l'y_{l-1}' \ldots y_2' y_1' \quad \textrm{and} \quad \omega' = a_k' a_{k-1}' \ldots a_2' a_1' .
$$

Let $\mathcal{U}$ be the monoid obtained according to Definition \ref{WWW} and let
$$
u =\langle \alpha ,  \omega \rangle \in \mathcal{U} .
$$

Then \emph{by definition}
$$
u' =\langle \alpha ' ,\omega ' \rangle .
$$

 Let $$\Gamma =\langle \mathcal{N}, \Sigma ,\Pi \rangle$$ be a grammar in Chomsky normal form. We construct the transition diagram $$ H_\Gamma =\langle V, R,\mathcal{U},l\rangle$$ obtained according to Definition \ref{H_Gamma}.

We will assume that every nonterminal symbol in $\Gamma$ is essential and therefore every vertex in $H_\Gamma$ is essential.

Let $G \subseteq \Sigma^* =(X\cup X' )^*$ be a group with decidable word problem, the set of generators $\Sigma =X\cup X'$, identity $\varepsilon$ (the empty word) and the set of defining  relations $\Theta$ such that
$$
\left\{ x_i x_i' =  x_i' x_i  =\varepsilon \; |\; i=1,2,\ldots ,m \right\} \subseteq \Theta .
$$

The next theorem is a direct consequence of Theorem \ref{th2} and Theorem \ref{Chomskyth}:

\begin{theorem}\label{thth6}

With the above notation let $A_i \in \mathcal{N}$, $i=1,2,\ldots , n$. Then the word $\alpha \in L(\Gamma ,A_i)$ and $\alpha \in \mathfrak{L} (G)$  if an only if there is a walk $\pi$ in the transition diagram $H_\Gamma$ with begin vertex $A_i$, end vertex $Z$ ($Z\notin \mathcal{N}$) and having label $l(\pi )=\langle \alpha , \omega \rangle =1_\mathcal{U}$, i.e. $\alpha =\varepsilon$ in the group $G$ and $\omega \in \mathfrak{D}_{2n}$, where $\mathfrak{D}_{2n}$ is restricted Dyck language on the $2n$ letters from $\mathcal{N}\cup \mathcal{N}'$, $n=|\mathcal{N} |=|\mathcal{N}' |$.

\hfill $\Box$
\end{theorem}

We consider the  semiring
$$
 \mathcal{S}_\mathcal{U} =\left( 2^\mathcal{U} ,\cup , \cdot ,\emptyset ,\{ 1_\mathcal{U} \} \right) ,
$$
 where $2^\mathcal{U}$ is the set of all subsets of $\mathcal{U}$. Operations in $\mathcal{S}_\mathcal{U}$ are respectively the union $\cup$ of sets and if $M_1,M_2\in 2^\mathcal{U}$ then by definition $M_1 \cdot M_2=M_1 M_2 =\{u\circ v\in \mathcal{U}\; |\; u\in M_1 , v\in M_2 \}$ (see equation (\ref{circ})), the zero is the empty set $\emptyset$ and the identity is the set $\{ 1_\mathcal{U} \} $  that contains only the identity $1_\mathcal{U} = \langle \varepsilon ,e \rangle $ of the monoid $\mathcal{U}$ (according to equation (\ref{unitofmonU})).

 In $H_\Gamma$, \emph{by definition}, we put
 $$
 A_{n+1} = Z ,\ \mathrm{i.e.} \ V=\left\{ A_1 ,A_2 ,\ldots ,A_n ,A_{n+1} \right\}  = \mathcal{N} \cup \left\{ A_{n+1} \right\} .
 $$

 We consider the following sets of walks in $ H_\Gamma$:
\begin{description}
  \item[$P_{ij}$] -- the set of all walks  $\pi\in  H_\Gamma$ with the initial vertex $A_i \in V$ and the final vertex  $A_j \in V$, $1\le i,j\le n+1$;

   \item[$\widehat{P_{ij}}$] -- the set of all walks $\pi\in  H_\Gamma$ with the initial vertex $A_i \in V$, the final vertex  $A_j \in V$, $1\le i,j\le n+1$, and in which all vertices are distinct, except possibly $A_i = A_j$. $\widehat{P_{ij}} \subseteq P_{ij}$;

  \item[$P_{iZ}$] -- the set of all walks  $\pi\in  H_\Gamma$ with the initial vertex $A_i \in V$ and the final vertex $Z=A_{n+1}$, $1\le i\le n+1$. ${P_{iZ}} \subseteq P_{ij}$;

  \item[$\widehat{P_{iZ}}$] -- the set of all walks  $\pi\in  H_\Gamma$ with the initial vertex $A_i \in V$, the final vertex $Z=A_{n+1}$, $1\le i\le n+1$, and in which all vertices are distinct, except possibly the initial and final vertices. $\widehat{P_{iZ}} \subseteq P_{iZ}$ as well as  $\widehat{P_{iZ}} \subseteq \widehat{P_{ij}}$;

  \item[$O_i $] -- the set of  all walks  $\pi\in  H_\Gamma$ with the initial vertex and the final vertex $A_i \in V$, $1\le i\le n+1$, and in which all vertices are distinct, except initial and final vertices which are $A_i$. $O_i =\widehat{P_{ii}}$.
\end{description}

Obviously
\begin{equation} \label{hhuhhuhhu}
L=L(\Gamma ,A_i )\subseteq  \mathfrak{L} (G) \Longleftrightarrow l (P_{iZ})=\{ 1_\mathcal{U} \} = \{ \langle \varepsilon ,e \rangle\} , \quad i=1,2,\ldots n.
\end{equation}

We consider the next elements of the semiring  $S_\mathcal{U}$:
\begin{description}
\item[$\Omega_3$] = $\displaystyle   \left\{ l (\pi ) \; \left| \; \pi \in \widehat{P_{1Z}} \right. \right\} =   l \left( \widehat{P_{1Z}} \right) $;
\item[$\Omega_4$] = $\displaystyle  \left\{ w'vw \; \left| \; \exists j\in  \{ 1,2,\ldots ,n+1\} \ :\  \exists \pi_1 \in P_{1j} ,\ v \in l(O_j ),\ w \in l(\widehat{P_{jZ}} ) \right. \right\} $;
\item[$W_2$] = $\displaystyle  \Omega_3 \cup \Omega_4  \in S_\mathcal{U} $.
\end{description}

We define the sets of walks $\mathcal{K}_{ij}^k$ in $H_\Gamma$, where $i,j\in\{ 1,2,..., n+1\}$, $k\in\{ 0,1,2,..., n+1\}$ $n = \vert \mathcal{N}\vert$ as follows:
$$
\mathcal{K}_{ij}^0 =
\left\{
\begin{array}{l}
\left\{ \rho \; \vert \; \rho =\langle A_i ,A_j \rangle {\rm \ is\ an\ arc\ in}\ R \right\}\quad \textrm{if}\ j\ne i\\
\left\{ \rho \; \vert \; \rho =\langle A_i ,A_i \rangle {\rm \ is\ a\ loop\ in}\ R \right\}\quad \textrm{if}\ j=i
\end{array}
\right.
$$
and
$$
\mathcal{K}_{ij}^k = \mathcal{K}_{ij}^{k-1} \cup \mathcal{K}_{ik}^{k-1} \mathcal{K}_{kj}^{k-1} .
$$

By definition $\mathcal{K}_{ij}^k$ consists only of walks  with  the initial vertex $A_i \in V$ the final vertex $A_j \in V$, and may not pass through a vertex $A_s$ when $s\ge k$, or that passes along a walk $\pi_1$ from $A_i$ to $A_k$, then passes along a walk $\pi_2$ from $A_k$ to $A_j$. None of these walks $\pi_1$ or $\pi_2$
passes along an interior vertex $A_s$ where $s \ge k$. So, for all $k\in\{ 0,1,\ldots ,n+1\}$ none of the walks $\pi\in \mathcal{K}_{ij}^k$
passes along an interior vertex $A_s$ where $s \geq k+1$.

\begin{proposition}\label{finitKiju}
The sets $\mathcal{K}_{ij}^k$ are finite.
\end{proposition}
Proof. By induction, it is easy to see that if $\pi \in \mathcal{K}_{ij}^k$, $i,j=1,2,\ldots ,n+1$, $k=0,1,\ldots ,n+1$ then the length of $\pi$ is less than or equal to  $2^k$, i.e every path $\pi \in \mathcal{K}_{ij}^k$ has a finite length. Therefore the sets $\mathcal{K}_{ij}^k$ are finite.

\hfill $\Box$

We consider the following elements of the semiring $\mathcal{S}_\mathcal{U}$:
\begin{description}
\item [$\Omega_5$] = $\displaystyle \left\{l (\pi ) \; |\; \pi\in \mathcal{K}_{1,\, n+1}^{n+1} \right\} = l \left( \mathcal{K}_{1,\, n+1}^{n+1} \right) $;
\item[$\Omega_6$] = $\displaystyle  \left\{ w'vw \; \left| \; \exists j \in  \{1,2,\ldots ,n+1\} \ :\ \exists \pi_1 \in \mathcal{K}_{1j}^{n+1} ,\ v \in l\left(\mathcal{K}_{jj}^{n+1} \right) ,\ w\in \ \left(\mathcal{K}_{j,\, n+1}^{n+1} \right) \right. \right\} $;
\item[$W_3$] = $\displaystyle  \Omega_5 \cup \Omega_6  \in \mathcal{S}_\mathcal{U} $.
\end{description}

It is not difficult to see that
\begin{equation}\label{OOOOmega}
\Omega_3 \subseteq \Omega_5 \subseteq l(P_{1Z} ) \quad \textrm{and} \quad \Omega_4 \subseteq \Omega_6 .
\end{equation}
Therefore
\begin{equation}\label{OOOOmega28}
W_2 \subseteq W_3
\end{equation}

As in $\mathcal{K}_{ij}^k$ is possible existence of a walk containing a cycle or a loop, then in the general case
$\Omega_3 \ne \Omega_5 \quad \textrm{and} \quad \Omega_4 \ne \Omega_6$.

\begin{theorem}\label{th2th}
Let $\Gamma =\langle \mathcal N, \Sigma ,\Pi \rangle$ be a grammar in Chomsky normal form, and let $L=L(\Gamma ,A_1)$, where $\mathcal{N} = \{A_1 ,A_2,\ldots ,A_n \}$, $\Sigma=X \cup X ' = \left\{ x_1 ,x_2,...,x_n \right\} \cup \left\{x_1 ' ,x_2 ' ,...,x_n ' \right\} $, $X \cap X ' =\emptyset$. Let $G$ be a group with the set of generators $\Sigma$, and the set of defining  relations $\Theta$ satisfying the condition (\ref{ThetaGroup}).

Then with the above notations and definitions, the following conditions are equivalent:

(i)  $L \subseteq \mathfrak{L} (G)$ ;

(ii)   $W_1 =\Omega_1 \cup \Omega_2 =\{ \varepsilon \}$ ;

(iii)   $W_2 =\Omega_3 \cup \Omega_4 =\{ 1_\mathcal{U} \} =\{ \langle \varepsilon , e\rangle \}$ ;

(iv)   $W_3 =\Omega_5 \cup \Omega_6 =\{ 1_\mathcal{U} \} =\{ \langle \varepsilon , e\rangle \}$ .

\end{theorem}

Proof. The equivalence of conditions (i) and (ii) was proved by A.V. Anisimov in \cite{springerlink:10.1007/BF01068773} (Theorem \ref{anisimov}).

Condition (i) $\Rightarrow$ (iv)  follows from the equations (\ref{hhuhhuhhu}) and (\ref{OOOOmega}).

Condition (iv)  $\Rightarrow$ (iii) follows from the equation (\ref{OOOOmega28}).

To prove the theorem, it remains to prove the condition (iii) $\Rightarrow$ (i).

Let $W_2 =\Omega_3 \cup \Omega_4 =\{ 1_\mathcal{U} \} = \{ \langle \varepsilon ,e\rangle \}$    and let    $\alpha \in L$. Then there is a walk   $\pi \in P_{1Z}$  such that  $l(\pi )=\langle \alpha ,e\rangle \in \mathcal{U}$.

If $\pi$ does not contain cycles and loops, then $l (\pi ) \in l (\widehat{P_{1Z}} ) =\Omega_3 =\{ 1_\mathcal{U} \} =\{ \langle \varepsilon  , e\rangle \}$  and therefore $\alpha =\varepsilon$ in $G$, i.e. $\alpha \in \mathfrak{L} (G)$.

If $\pi$ contains a cycle or a loop  then there is $A_j \in V$ such that $\pi$ can be expressed as $\pi =\pi_1 \pi_2 \pi_3$, where $\pi_1 \in P_{1j} , \pi_2 \in O_j , \pi_3 \in \widehat{P_{jZ}}$ and $(l (\pi_3 ))' l (\pi_2 )l (\pi_3 ) \in \Omega_4 =\{ 1_\mathcal{U} \}$. Therefore, $l (\pi_2 )l (\pi_3 )=l (\pi_3 )$ and $l (\pi_1 \pi_2 \pi_3 )=l (\pi_1 \pi_3 )$. Since $\pi_2 \in O_j$, then the length of $\pi_2$ is greater than 1. Consequently, in $H_\Gamma$ there is a walk with length less than the length of $\pi$, whose label is equal to $l(\pi )=\langle \alpha ,e\rangle$ in the semiring $\mathcal{U}$. This process of reduction may proceed a finite number of times as the length of $\pi$ is finite. At the end of this process we obtain a walk in $H_\Gamma$ with the initial vertex $A_1$ and the final vertex  $A_Z =A_{n+1}$ without cycles and without loops with label equal to $\langle \alpha , e\rangle \in\mathcal{U}$. But $l (\widehat{P_{1Z}} )=\Omega_3 =\{ 1_\mathcal{U} \} =\{ \langle \varepsilon ,e \rangle \}$. Hence $\alpha =\varepsilon$ in the group $G$ and therefore $L \subseteq \mathfrak{L} (G)$.

\hfill $\Box$

Let $M_1,M_2\in 2^\mathcal{U}$. In the semiring $\mathcal{S}_\mathcal{U}$ we define the next binary operation:
 \begin{equation}\label{[x,y]}
M_1 \star M_2 =\left\{ w' vw \; | \; v\in M_1 , w\in M_2 \right\}
 \end{equation}

The following algorithm is based on the equivalence (i) and (iv) of Theorem \ref{th2th}.  For convenience,  $g_{ij}^k$, $i,j,k\in \{1,2,\ldots , n+1\}$  will mean $l (\mathcal{K}_{ij}^k )$. Here, $k$ in $g_{ij}^k$ is a superscript and does not mean an exponent.

\begin{algorithm}\label{alg1}
Verifies the inclusion $L \subseteq \mathfrak{L} (G)$ for a regular language $L$, and a group language $\mathfrak{L} (G)$, where $G$ is a group with  decidable word problem.
\end{algorithm}

{\bf Input:} $g_{ij}^0 =l (\mathcal{K}_{ij}^0 ),\quad i,j=1,2,...,n+1$

{\bf Output:} Boolean variable $T$, which receives the value {\bf True} if $L \subseteq \mathfrak{L} (G)$, and the value {\bf False}, otherwise. The algorithm will stop immediately after the value of $T: = \textbf{False}$.

Begin

1. $T:= \textbf{True} $;

2. For  $1\le k\le n+1$       Do

3. \hspace{0.4cm}  For $1\le i,j\le n+1$  Do

4. \hspace{0.8cm} $g_{ij}^k :=g_{ij}^{k-1} \cup g_{ik}^{k-1} g_{kj}^{k-1}$;

5. \hspace{2.8cm}  End Do;

6. \hspace{2.0cm}  End Do;

7. If $g_{1\,n+1}^{n+1} \ne \emptyset$  and $g_{1\, n+1}^{n+1} \ne \{ \langle \varepsilon ,e\rangle \}$   Then

8. \hspace{2.4cm}  Begin  $T:=\textbf{False}$; Halt; End;

9. For $1\le j\le n+1$   Do

10.\hspace{0.4cm} If $g_{1j}^{n+1} \ne \emptyset$ and $g_{jj}^{n+1} \ne \emptyset$ and $g_{j\, n+1}^{n+1} \ne \emptyset$  Then

11.\hspace{0.8cm} If $g_{jj}^{n+1} \star g_{j\, n+1}^{n+1} \ne \{ \langle \varepsilon ,e\rangle \}$  Then

12.\hspace{2.5cm}  Begin  $T:=\textbf{False}$; Halt; End;

13.\hspace{2cm}  End Do;

End.

\begin{theorem}\label{tttt3}
Algorithm \ref{alg1} checks the inclusion $L\subseteq \mathfrak{L} (G)$, where $L$ is a context-free language generated by a grammar in Chomsky normal form  with $n$ nonterminals, $\mathfrak{L} (G)$ is a group language, which specifies the group $G$ with  decidable word problem. Algorithm \ref{alg1} executes at most  $O(n^3 )$ operations $ \cup $ and $\cdot$, and at most $O(n^2 )$ operations $ \star $ in the semiring $\mathcal{S}_\mathcal{U}$, where the binary operation  $ \star $ is defined using the equation (\ref{[x,y]}).
\end{theorem}

Proof.  According to Theorem \ref{th2th} and considering axioms of the semiring  $\mathcal{S}_\mathcal{U}$, then in rows 8 and 12 of Algorithm \ref{alg1}, the boolean variable $T$ gets the value \textbf{False} if and only if $L$ is not included in $\mathfrak{L} (G)$. Otherwise, $T$ gets the value \textbf{True}. Hence the algorithm correctly checks whether the inclusion $L\subseteq \mathfrak{L} (G)$ is true.

 It is easy to see that line 4  executes no more than $(n+1)^3$ times. During each iteration, the operations $\cup$ and $\cdot$ perform in the semiring  $\mathcal{S}_\mathcal{U}$ once each of them. Lines 10 and 11 is executed at most $(n+1)^2$ times each. Therefore, Algorithm \ref{alg1} performs no more than $O(n^3 )$ operations  $ \cup $ and $ \cdot $, and no more than $O(n^2 )$ operations $\star $  in the semiring $\mathcal{S}_\mathcal{U}$. The theorem is proved.
\hfill $\Box$

\begin{corollary}
If the operations  $\cup $, $\cdot $ and $\star $ in the semiring $\mathcal{S}_\mathcal{U}$  can be done in a polynomial time, then Algorithm \ref{alg1} is polynomial.

\hfill $\Box$
\end{corollary}

\begin{remark}
 According to Proposition \ref{finitKiju} the sets $l(\mathcal{K}_{ij}^k )\in \mathcal{S}_\mathcal{U}$ are finite and  hence they can be coded using a Boolean vector of finite length. In this case, to evaluate algorithm \ref{alg1} more accurately, it is convenient to use bitwise operations \cite{Yordzhev2019,RN110}.
\end{remark}


\end{document}